\documentclass{ecai}
\usepackage{graphicx}
\usepackage{latexsym}

\usepackage{algorithm} 
\usepackage{microtype}
\usepackage{subfigure}
\usepackage{booktabs} 
\usepackage{mathtools}
\usepackage{amsmath}
\usepackage{verbatim} 
\usepackage{algpseudocode}
\usepackage{multirow}
\usepackage{xcolor,colortbl}

\definecolor{green}{rgb}{0.67, 0.91, 0}
\definecolor{red}{rgb}{1.01, 0.38, 0.20}

\ecaisubmission   

\begin{document}

\begin{frontmatter}

\title{Mapping Global Value Chains at the Product Level}

\author[A]{\fnms{Lea}~\snm{Karbevska}\orcid{0000-0003-3712-2266}\thanks{Lea Karbevska. Email: karbevska.lea@gmail.com.}}
\author[A,B,C]{\fnms{C\'esar A.}~\snm{Hidalgo}\orcid{0000-0002-6977-9492}\thanks{C\'esar A. Hidalgo. Email: cesifoti@gmail.com.}}

\address[A]{Center for Collective Learning, ANITI, TSE-R, IAST, IRIT, Université de Toulouse, 31000 Toulouse, France}
\address[B]{Alliance Manchester Business School, University of Manchester, United Kingdom.}
\address[C]{Center for Collective Learning, CIAS, Corvinus University, Budapest, Hungary.}

\begin{abstract}
Value chain data is crucial to navigate economic disruptions, such as those caused by the COVID-19 pandemic and the war in Ukraine. Yet, despite its importance, publicly available value chain datasets, such as the ``World Input-Output Database'', ``Inter-Country Input-Output Tables'', ``EXIOBASE'' or the ``EORA'', lack detailed information about products (e.g. Radio Receivers, Telephones, Electrical Capacitors, LCDs, etc.) and rely instead on more aggregate industrial sectors (e.g. Electrical Equipment, Telecommunications). Here, we introduce a method based on machine learning and trade theory to infer product-level value chain relationships from fine-grained international trade data. We apply our method to data summarizing the exports and imports of 300+ world regions (e.g. states in the U.S., prefectures in Japan, etc.) and 1200+ products to infer value chain information implicit in their trade patterns. Furthermore, we use proportional allocation to assign the trade flow between regions and countries. This work provides an approximate method to map value chain data at the product level with a relevant trade flow, that should be of interest to people working in logistics, trade, and sustainable development.
\end{abstract}

\end{frontmatter}

\section{Introduction}
Value chain data is important to understand the resilience and systemic effects of disruptions, such as natural disasters \cite{PARK201375,doi:10.1177/0972150913501606}, climate change \cite{doi:10.1080/00207543.2019.1629670}, war \cite{ukraine_war,RePEc:ocp:ppaper:pb11-22}, and disease \cite{/content/publication/2a7081d8-en}. Publicly available value chain data, such as the OECD Inter-Country Input-Output Database \cite{OECD2021}, the World Input-Output Database \cite{https://doi.org/10.1111/roie.12178}, EXIOBASE \cite{stadler2018exiobase}, and EORA \cite{doi:10.1080/09535314.2013.769938,doi:10.1021/es300171x}, however, have limited sectoral resolution, and are often disaggregated into a few dozen industries. This high-level of aggregation can be limiting \cite{doi:10.1146/annurev-economics-080217-053600, RePEc:arx:papers:2302.11451} for applications where detailed product or sectoral resolution is needed, such as tracing the environmental impact and social responsibility of suppliers.\\

On the contrary, international trade data is much more disaggregated, with 5,000+ categories at the ``six-digit level'' (HS6 \cite{chaplin1987introduction}) and 1,000+ categories at the ``four-digit level'' (HS4). Yet, while international trade data is also a go-to dataset for analysts working to understand disruptions, trade data lacks explicit information about value chain relationships. Trade data can tell us that China imports iron ore from Brazil, but it cannot tell us what that iron ore is used for (e.g. cars, iron rods, aircraft, etc.). Nevertheless, trade theory tells us that this data must contain implicit information about value chain relationships. This information should be hidden in a country or region's specialization patterns and we should be able to extract it by combining trade theory inspired features with machine learning techniques.\\

The idea of mapping value chains, however, is not new. Several projects have tried to combine input-output tables \cite{leontief1986input} and trade data in efforts to map global value chains \cite{doi:10.1080/09535314.2013.769938,doi:10.1021/es300171x, https://doi.org/10.1111/roie.12178, OECD2021}. These efforts use national input-output tables, connecting sectors at the local level with trade data, to estimate the volume of imported inputs used in each sector of an economy. These efforts, however, tend to rely on proportional allocation methods, where imports are distributed among sectors in the same proportion as local inputs. That is, they assume, for instance, that if 20\% of the steel produced in a country is used for the production of machinery, then 20\% of the steel imported from any country will also be used for the production of machinery. The result is several datasets \cite{doi:10.1080/09535314.2013.769938,doi:10.1021/es300171x, https://doi.org/10.1111/roie.12178, OECD2021}, that while useful and important, could benefit from better sectoral and spatial resolution. \\  

There is consensus that detailed value chain data can be crucial for a number of applications. Consider the disruptions caused by the Evergreen, the massive container ship that in 2021 became stranded in the Suez Canal \cite{noauthor_suez_nodate_bbc}. By blocking the Suez Canal, the Evergreen impeded the flow of a large number of products between Asia and Europe \cite{noauthor_suez_nodate_dw}. Value chain data can also inform questions with geopolitical implications. Often countries prefer to avoid sourcing key components from geopolitical rivals. For instance, they try to organize their value chains in ways that avoid depending on potential enemies for strategic resources, such as fuel and electronics. Moreover, value chain data can be a key input for environmental assessments \cite{stadler2018exiobase}, since it is needed to account for the environmental impact of imported goods. Finally, value chain data can be important to those working on corporate responsibility. For instance, a clothing company may want to have traceability of their inputs to ensure its products are not produced using forced or child labour.\\

Yet, despite the glaring need for value chain data, there are no detailed publicly available value chain datasets with fine spatial and sectoral resolution. In this paper, we explore the creation of a method to infer value chain relationships from international trade data to create high-resolution maps of global value chains.\\

Our method exploits the idea that geographies that specialize in the export of a certain product will tend to specialize in the procurement of its inputs. This tendency should be observed twice in trade data: upstream and downstream. The upstream tendency should be expressed in the products imported by a location that specializes in the export of a product. That is, we expect exporters of computers to specialize in the import of LCDs. Similarly, the downstream tendency should be expressed in the products exported by a location specialized in the import of a product. That is, importers of LCDs will tend to specialize in the export of computers. Here we combine both upstream and downstream specialization patterns in a model that we optimize to identify input-output relationships. We apply this model to a dataset summarizing the exports and imports of 300+ world regions (e.g. states in the U.S., prefectures in Japan, etc.) to create a product-level dataset of value chain relationships for 1,200+ products.\\ 

Our method, however, is not without its limitations. While it is designed to operate at the product level, it is not perfectly accurate, meaning that it provides some false-positive value chain relationships. Also, it does not provide a full input-output network, but a set of the most likely value chain links for each product. Moreover, our method requires optimizing four different parameters, a process that can be slow and complicated. Despite these limitations, our results show some promise by identifying 100s of value chain links at the product level. This validates the possibility of using international trade data at the regional level to identify value chain relationships.\\

In the remaining sections of this paper, we provide a detailed description of our data and methods. The structure of the paper is as follows: we begin with a section that discusses the data utilized in our model (section \emph{Data}). Following this, we introduce some of the essential trade theory concepts that underpin our approach (section \emph{Trade Theory}). Then we illustrate a model that uses proportional allocation to assign trade flow in a value chain and, consequently, demonstrating that the missing part of the model is the input-output relationship between products (section \emph{Conceptual Model}). Next, we introduce and explain our ``Backward \& Forward" method which predicts input-output relationships between products (section \emph{Methodology}). Furthermore, we use the ``Backward \& Forward" method to construct a product-level dataset fine-tuned on the OECD Inter-Country Input-Output table \cite{OECD2021}, and we validate the results obtained. Our findings are expected to contribute significantly to the development of computational methods aimed at constructing global value chain datasets.

\section{Data}

We leverage fine-grained international data compiled by the Observatory of Economic Complexity \cite{Simoes2011TheEC} (oec.world) spanning from the year 2017 to 2020. This is data on exports and imports at the regional level for 1,226 HS4 level and 5,890 HS6 level unique products. Because of incompatibilities in data reporting (not all countries report regional trade data using the same classification), our sample is limited to 9 countries: Brazil (32 regions), Canada (13 regions), Chile (16 regions), China (31 regions), Japan (41 regions), Mexico (32 regions), Russia (85 regions), Spain (53 regions) and the United States of America (54 regions).
    

We clean this dataset by removing  unknown regions and reexports such as ``Reexportação'', ``Exterior'', ``Mercadoria Nacionalizada'' and ``Consumo de Bordo'' from the data of Brazil, ``Unknown'' from the data of the USA and ``Sin provincia asignada'' from the data of Spain. This leaves us with 351 regions.\\

\begin{figure}
\begin{center}
     {\includegraphics[width=\columnwidth]{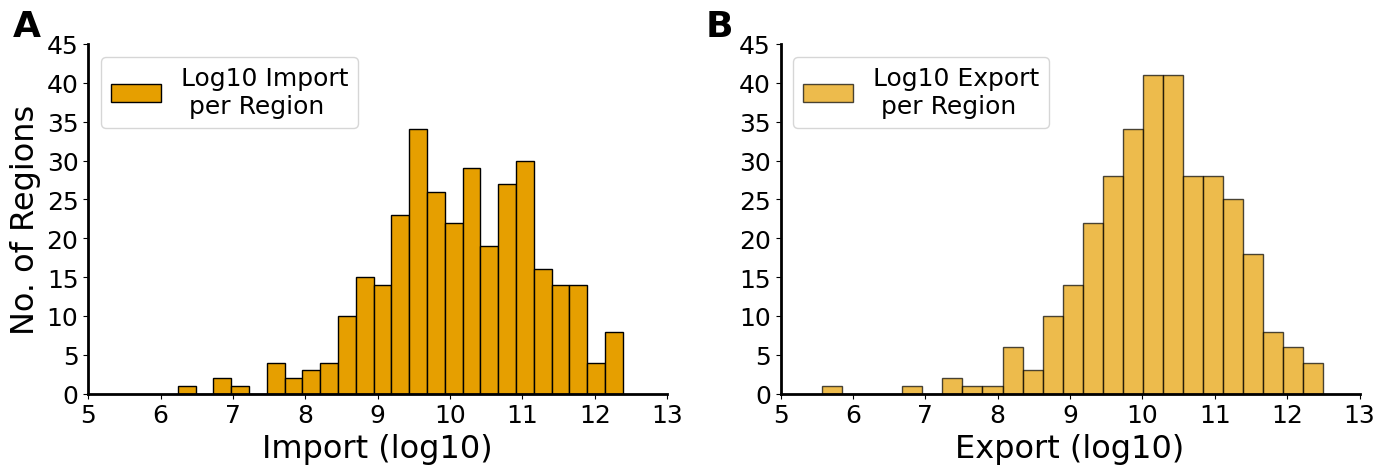}}
     \caption{Total import (A) and export (B) distributions across 9 regions from 2017-2020 from the Observatory of Economic Complexity data. Regions include Brazil, Canada, Chile, China, Japan, Mexico, Russia, Spain, and USA.}\label{fig:dist_imp_exp}
\end{center}
\end{figure}

We then remove small regions that can provide a noisy signal about their exports and imports (a few dollars of exports and imports can drastically change the observed specialization pattern of regions with low trade volumes). After inspecting the distribution of exports and imports (Figure \ref{fig:dist_imp_exp}), aggregated from 2017 to 2020, we remove regions on the left tails of the exports and imports distribution. These are regions which in these four years imported in total less than 100 million USD and exported in total less than 1 billion USD. After removing these 44 regions we are left with a final sample of 306 regions.

We note that our data contains export and import information between regions and countries. That is, we know what Barcelona imports from Brazil, or what Sao Paulo imports from Spain, but not what is traded between Barcelona and Sao Paulo. \\

Finally, we reconcile exports and imports (look for the products to be available in both the exports and import records) and are left with a total of 1,220 HS4 and 5,890 HS6 unique product categories.\\

In addition, we use the 2021 edition of OECD  Inter-Country Input-Output data to fine-tune our model. This is a table containing 45 unique industries based on ISIC Revision 4 (industry, not product categories) for 66 countries. From this data, we produce two tables: OECD specialization and the OECD labeled data. A description of this data can be found at: \cite{OECD2021}.
    
\section{Trade Theory}

Trade theory is the branch of economics studying the patterns in regional and international trade. It has a mathematical tradition of over 200 years, starting with the works of David Ricardo \cite{ricardo2005principles}. In this paper, we use trade theory to create some of the basic features for our model.\\

A key concept in trade theory is the idea of \emph{comparative advantage}. A location is said to have \emph{comparative advantage} in the products that it is specialized in. Trade theory tells us that \emph{comparative advantages} should tell us about the factors that an economy is well endowed with. For instance, we expect economies endowed with vast maritime resources to specialize in the exports of fish and landlocked mountainous economies to specialize in the exports of minerals.\\

In today's globalized economy, however, where intermediate inputs are highly mobile, economies often specialize in processes that are not necessarily pinned down by the presence of natural resources but by the availability of knowledge \cite{10.2307/j.ctt9qf8jp}. That is, countries that export cars or furniture do not do so because they are endowed with vast reserves of iron or lumber (they can source these from global markets). This means that countries and regions will tend to import some of the inputs they need to produce the outputs they export. Thus, we should be able to observe value chains implicitly, albeit imperfectly, in international trade flows.\\

Estimating \emph{comparative advantages} in practice, however, can be challenging because countries and products can vary greatly in size. Trade theory scholars use indicators of Revealed Comparative Advantage (RCA) \cite{https://doi.org/10.1111/j.1467-9957.1965.tb00050.x} (also known as the Location Quotient in urban economics), to measure the level of specialization of a location in a product.\\

Formally, the Revealed Comparative Advantage of a location in an activity (e.g. a country or region in a product) is simply the double normalization of the export matrix. That is, the $RCA$ of a location $l$ in a product $p$ is defined as:
    \begin{align}
        RCA_{lp} &= \frac{X_{lp}}{ \sum_{p'} X_{lp'}} / \frac{\sum_{l'} X_{l'p}}{\sum_{l'p'} X_{l'p'}},
        \label{RCA}
    \end{align}
where $X_{lp}$ are the exports of location $l$ in product $p$.\\

$RCA$ can also be interpreted as the ratio between observed and expected exports. When a location has an RCA larger than 1 in a product, we say that the location is specialized in that product since it exports more than what it is expected for a location of the same size and for a product with the same global market.\\

Going forward, we define two versions of $RCA$. An export $RCA^{export}$, as defined in equation \ref{RCA}, and an import $RCA^{import}$ defined in the same manner, but where $X_{lp}$ represents the imports of location $l$ in product $p$. The $RCA^{import}$ should tell us about the product that a region imports too much. Our hypothesis is that by exploiting specialization patterns across multiple geographies we can generate features that when fed in a machine learning model can recover information about global value chains.

\section{Conceptual Model}

Formally, our goal is to estimate the tensor $X_{r_1 p_1 r_2 p_2}$, representing the flow of product $p_1$ coming from region $r_1$ and used in region $r_2$ to produce product $p_2$. The data we have available, however, is much more incomplete and represents two aggregates of the aforementioned tensor. These are: $X_{r_1 p_1 c_2}$ and  $X_{c_1 p_1 r_2}$, which denote, respectively, the exports of region $r_1$ of product $p_1$ to country $c_2$ (where region $r_2$ is located) and the imports of region $r_2$ of product $p_1$ coming from country $c_1$ (where region $r_1$ is located).\\

\begin{figure*}
     \centering
     {\includegraphics[width=0.6\linewidth]{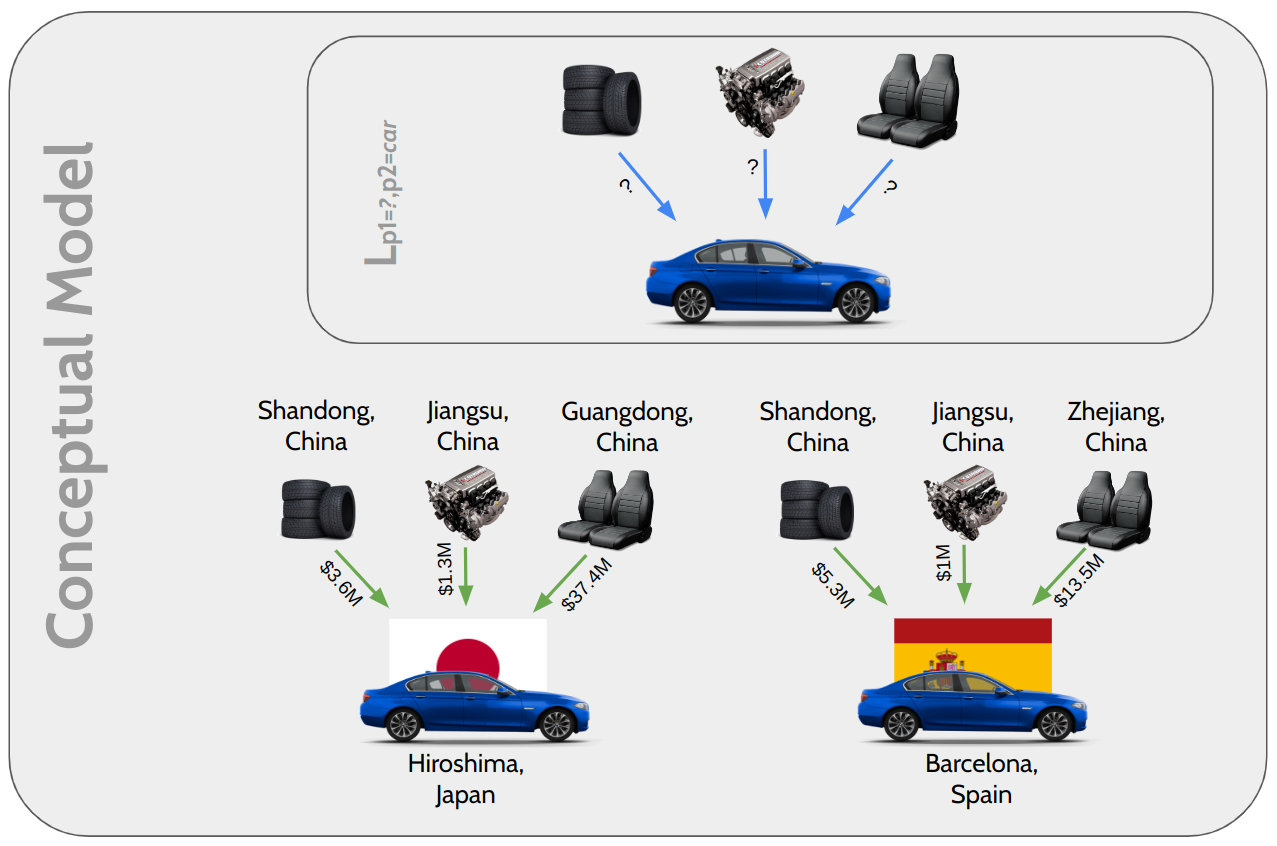}}
     \caption{The Conceptual Model utilizes the input-output relationship produced by the L term and further quantifies the trade flow between two regions. In this figure, the term L provides the information that Cars need Rubber Tires, Engine Parts and Seats for their production. Whereas, the Conceptual Model uses the input-output information to quantify the exports of Rubber Tires, Engine Parts and Seats from Hiroshima to Barcelona which would be used for Barcelona's local production of Cars.}\label{fig:general_model}
\end{figure*}

We estimate $X_{r_1 p_1 r_2 p_2}$ using the following proportional allocation model:

\vskip -0.2in
\begin{equation}
X_{r_1 p_1 r_2 p_2}=\frac{X_{r_1 p_1 c_2}}{\sum_{r_1}X_{r_1 p_1 c_2}} \underbrace{L_{p_1 p_2}}_{\text{unknown}} \frac{X_{r_2 p_2}}{\sum_{p_2} X_{r_2 p_2}}  X_{c_1 p_1 r_2},  
\label{general_model}
\end{equation}
where the first fraction represents the share of exports in region $r_1$ of product $p_1$ to country $c_2$ over the total exports of product $p_1$ to country $c_2$. This term shows us how specialized in export of product $p_1$ to country $c_2$ is the region $r_1$ in comparison with all the other regions from its country $c_1$. $L_{p_1 p_2}$ is a binary tensor where 1 represents a positive input-output relationship between the products $p_1$ and $p_2$ (Figure \ref{fig:general_model}). However, from our available data the $L$ matrix is not known. The second fraction is the share of export of product $p_2$ in region $r_2$ over all the exports of region $r_2$. This term shows us how specialized the region $r_2$ is in the export of the product $p_2$. Lastly, $X_{c_1 p_1 r_2}$ is the trade flow in exports of product $p_1$ from country $c_1$ to region $r_2$. 

Consider the example where we want to find the trade flow of \emph{engine parts} from \emph{Jiangsu Province} (region in China) to \emph{Barcelona} (region in Spain). In this example, $\frac{X_{r_1 p_1 c_2}}{\sum_{r_1}X_{r_1 p_1 c_2}}$ is the share of \emph{Jiangsu} in the export of Chinese \emph{engine parts} to Spain. The $L_{p_1 p_2} \frac{X_{r_2 p_2}}{\sum_{p_2} X_{r_2 p_2}}$ term would be the predicted output products of \emph{engine parts} (ex. cars, trucks, boats etc.) multiplied by the share of export of \emph{Barcelona} of those products. And $X_{c_1 p_1 r_2}$ would be the imports of Chinese \emph{engine parts} by \emph{Barcelona}. By multiplying the previously mentioned terms we should be able to estimate how much \emph{Barcelona} spends in imports of \emph{engine parts} from \emph{Jiangsu Province} for i local production of ex. cars, trucks, and boats.\\

The data that we have access to provides the information to calculate the term $X_{c_1 p_1 r_2}$, and the two fractions: $\frac{X_{r_1 p_1 c_2}}{\sum_{r_1}X_{r_1 p_1 c_2}}$ and  $\frac{X_{r_2 p_2}}{\sum_{p_2} X_{r_2 p_2}}$. However, the terms ${X_{r_1 p_1 r_2 p_2}}$ and $L_{p_1 p_2}$ are unknown.\\
In the next chapter, Methodology, we present a method to calculate the binary matrix $L_{p_1 p_2}$.
\section{Methodology}

\subsection{Backward \& Forward Method}
A link in a value chain can be traversed in two directions: a downstream or forward direction (from sunflower seeds to sunflower oil) and an upstream or backward direction (from sunflower oil to sunflower seeds) \cite{SINGER2008669}.\\ 

To estimate the term $L_{p_1 p_2}$ from Equation \ref{general_model} we introduce the ``Backward \& Forward'' method.
The ``Backward \& Forward'' method combines the ``Forward'' and the ``Backward'' approaches inspired by the downstream and upstream value chain flows.\\

In the ``Forward'' approach, we start by selecting an import product $C$ and then we select the regions that import a disproportionately large number of product $C$ (using an  $RCA^{import}$ measure). This provides us with a list of locations sorted by import RCA (e.g. Alabama, Aguascalientes, etc.), which are places that import ``too much'' of that product. We then look at the export specialization of these regions. The result is a matrix of the exports of the locations that import ``too much'' of a product. We then attempt to learn the outputs associated with the import from the over-expressed exports of these locations. 

In the ``Backward'' approach we first select an export product $P$ and then identify the locations that export a disproportionately large number of that product. We then analyse what these locations specialize in, in terms of imports. The result is a matrix of the imports of the locations that export ``too much'' of the selected product. We then attempt to learn the inputs of that product using the values from the matrix.\\


 Every product has an input but not every product has an output. For example, ``Rolled Tobacco'' (aka cigarette) is a final product that goes directly into consumption. While, raw materials as "Iron Ore" still need excavation machines to be extracted and transported. For that reason, we identify inputs of every product by first applying the ``Backward'' and then validating with the ``Forward'' approach. We call this the ``Backward \& Forward'' method.\\
    
In pseudo-code (Algorithm \ref{alg:back_forw_algo}), we identify the inputs of all of our products $P$ by fixing a product $P_i$ ($P_i\in P$ where $1\le i\le No.\_of\_products$) and apply the ``Backward'' approach first (``get\_n\_input\_candidates()''). This gives us the top n input candidates $C$ ($C_j\in C$ where $1\le j\le n$) for the input $P_i$. If we find $P_i$ as an input to itself in $C$ we remove it (``drop()''). Then, for each $C_j$, we apply the ``Forward'' approach (``get\_n\_output\_candidates()'') to identify the outputs of $C_j$, called $T$. We then look for product $P_i$ in the outputs ($T$) of $C$. If we find $P_i$ in $T$, we take the rank (``getRank()'') of $P_i$ in $T$ and add it to the rank of $C_j$ in the inputs of $P_i$. And if we do not find $P_i$ in $T$, then we take the worst ranking which is the one of the last $P_n$ candidate product and add plus one, and add it to the rank of $C_j$ in the inputs of $P_i$. This technique updates (``updateRank()'') the initial ordering of $C_j$ as an input of $P_i$. We then order the products by ascending order (``order\_by\_rank\_ascending()''). 
This makes ``2'' the minimum and best possible rank meaning that $C_j$ was the first candidate (rank 1) as an input to $P_i$ and $P_i$ was the first candidate (rank 1) as an output to $C_j$.\\
By merging both the ``Backward'' and then the ``Forward'' method, we are able to first identify and then validate and update our input-product results. 

\begin{algorithm}[tb]
    \caption{Backward \& Forward Algorithm}
    \small
    \begin{algorithmic}
    \For{\texttt{every $P_i \in P$}}
        \State \texttt{
            \State $C \gets get\_n\_input\_candidates(P_i)$ \Comment{\scriptsize Backward}
            \small
            \If{$P_i$ in $C$}
                \State $C.drop(P_i)$
            \EndIf
            \For{\texttt{every $C_j \in C$}}
               \State \texttt{
                \State $T \gets get\_n\_output\_candidates(C_j)$ \Comment{\scriptsize Forward}
                \small
                \If{$P_i \in T$}
                    \State $value \gets T.getRank(P_i)$
                \Else
                    \State $value \gets T.getRank(T_n)+1$
                \EndIf
                \State $old\_rank \gets C.getRank(C_j)$
                \State $new\_rank \gets old\_rank+value$
                \State $C.updateRank(C_j,new\_rank)$
                \Comment{\scriptsize Update ranking}
                \small
               }
            \EndFor\\
            \State $result \gets C.order\_by\_rank\_ascending()$
        }
    \EndFor
    \end{algorithmic}
    \label{alg:back_forw_algo}
    \end{algorithm}

\subsection{Fine-tuning}
    In the ``Backward \& Forward'' method we come across a few parameters that we need to optimize: two parameters in the ``Backward'' and two in the ``Forward'' approach.\\

    In the ``Backward'' approach we have the specialization of exporters threshold ($rca\_locations\_1$). This is the minimum $RCA^{export}$ a location needs to have to be considered specialized in the export of a certain product. Then, we have the import specialization of industries/products threshold ($rca\_industries\_1$). This helps us rank the input candidates by counting how many of the specialized export locations, that we identified before, are also specialized in the import of the input candidate.\\
    
    Similarly, in the ``Forward'' approach, we have the specialization of importers threshold ($rca\_locations\_2$). This is the minimum $RCA^{import}$ a location needs to have to be considered specialized in the import of a certain product. Then, we have the export specialization of industries/products threshold ($rca\_industries\_2$). This helps us rank the output candidates by counting how many of the specialized import locations, that we identified before, are also specialized in the export of the output candidate.\\

     To optimize the parameters $rca\_locations\_1$, $rca\_industries\_1$, $rca\_locations\_2$ and $rca\_industries\_2$ we feed the OECD ICIO specialization data to the ``Backward \& Forward" Model. Then we use the OECD labeled data to evaluate the model.\\
    
    Considering the nature of our value chain problem, the full result would be a sparse matrix with many 0s and very few 1s. Meaning that a big part of all the possible product pairs' input-output relationships (1488400 HS4 and 34692100 HS6 products) will be false and very few will be true. The model should prioritize identifying the true relationships. For this reason, we use precision as an evaluation metric which tells us how many of our predicted input-output relationships are correctly classified \cite{https://doi.org/10.48550/arxiv.2010.16061}.\\
 
\begin{figure}[ht]
    \begin{center}
\centerline{\includegraphics[width=0.95\columnwidth]{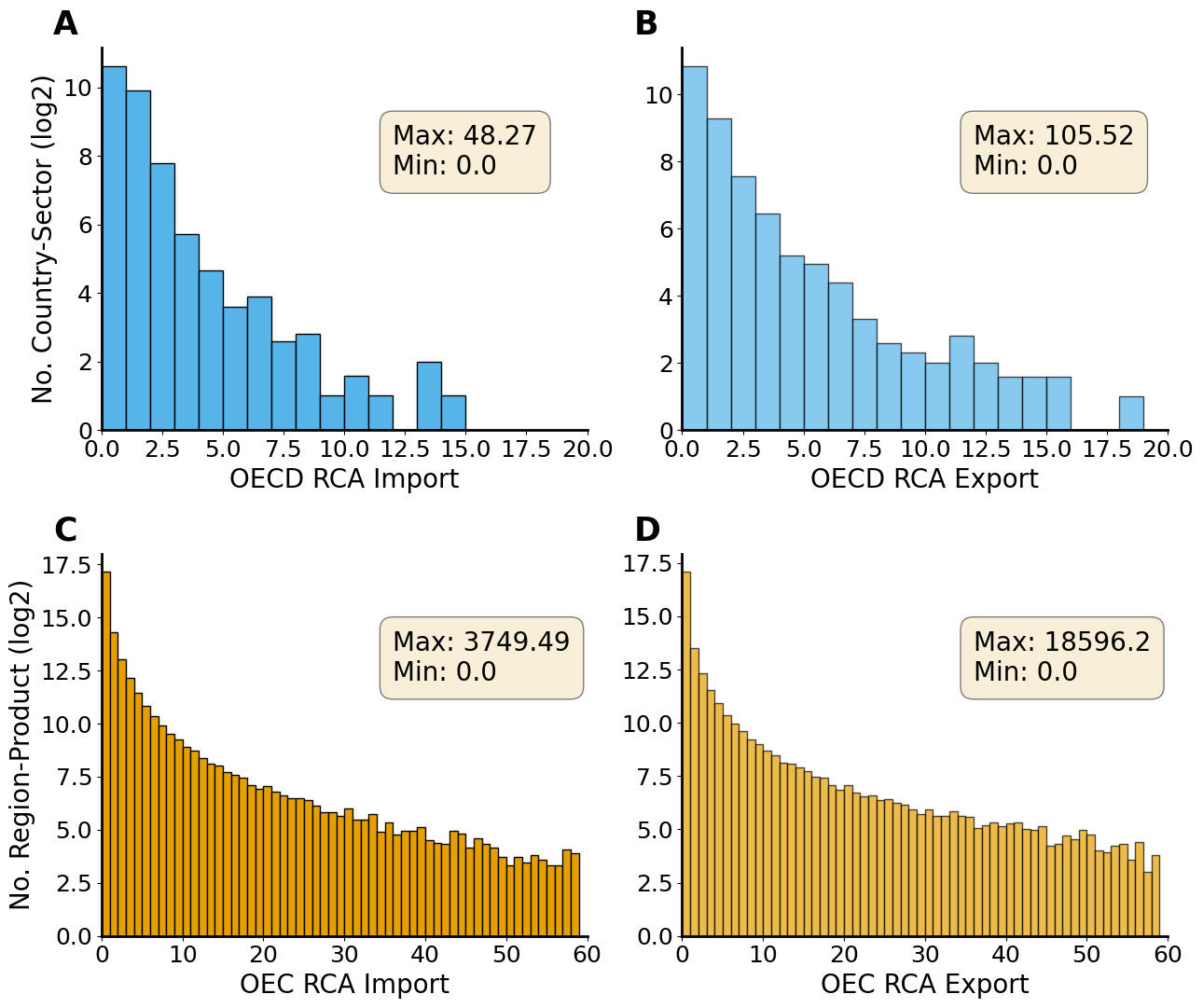}}
    \caption{Total OECD RCA import (A), OECD RCA export (B), OEC RCA import (C) and OEC RCA export (D)  distribution across all the product-region in the case of the OEC and the sector-country pairs in the case of OECD data between 2017 and 2020.}
    \label{fig:oecd_oec_rca_dist}
    \end{center}
    \end{figure}

    To optimize the parameters we perform a grid search. Looking at the distributions of RCAs in Figure \ref{fig:oecd_oec_rca_dist}, we test all the combinations of $rca\_locations\_1$, $rca\_industries\_1$, $rca\_locations\_2$ and $rca\_industries\_2$ assigned with a value in the interval of $\left[1,6\right[$ with step of 0.5. For every combination, we calculate the corresponding precision.

    \begin{figure*}
    \begin{center}
\centerline{\includegraphics[width=0.9\linewidth]{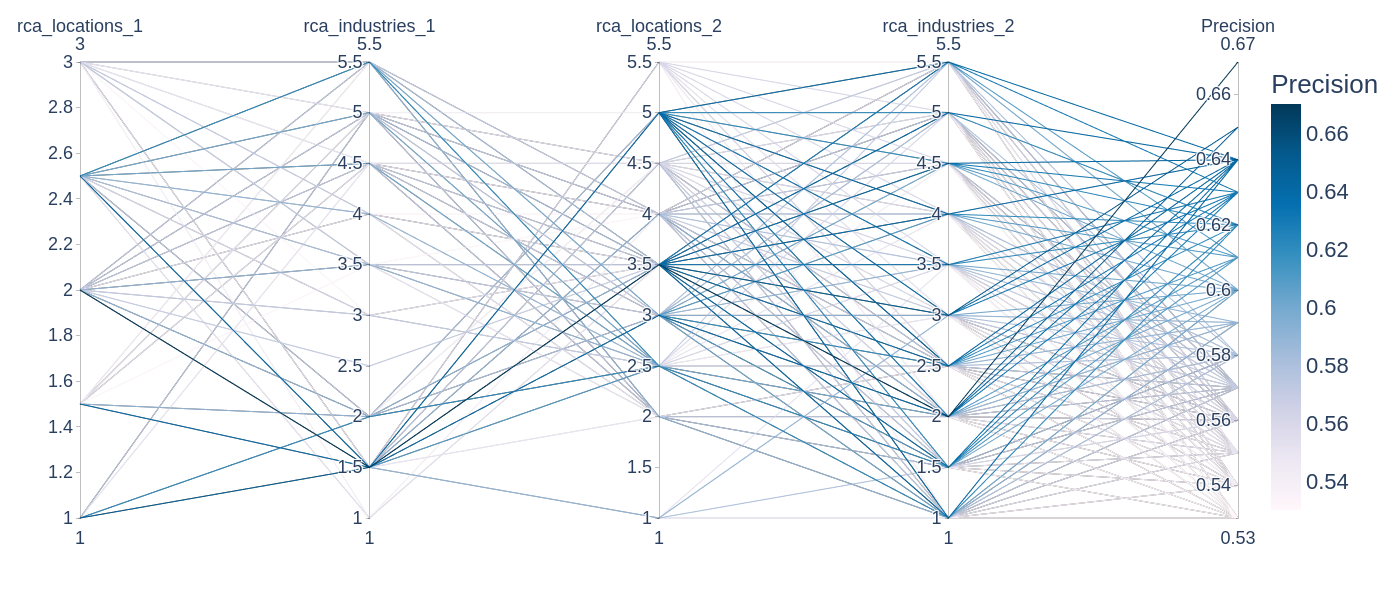}}
    \caption{Parallel coordinate plot showing the 500 highest in precision combinations of the parameters: rca\_location\_1, rca\_industries\_1, rca\_location\_2 and rca\_industries\_2. The precision values are rounded to two decimals.}
    \label{fig:training_everything}
    \end{center}
    \end{figure*}
    
    The Figure \ref{fig:training_everything} shows the 500 combinations that achieved the highest precision. The best precision is 0.67 with the $rca\_locations\_1=2$, $rca\_industries\_1=1.5$, $rca\_locations\_2=3.5$  and $rca\_industries\_2=2$.\\
    
\begin{figure}[h]
    \begin{center}
\centerline{\includegraphics[width=1.1\columnwidth]{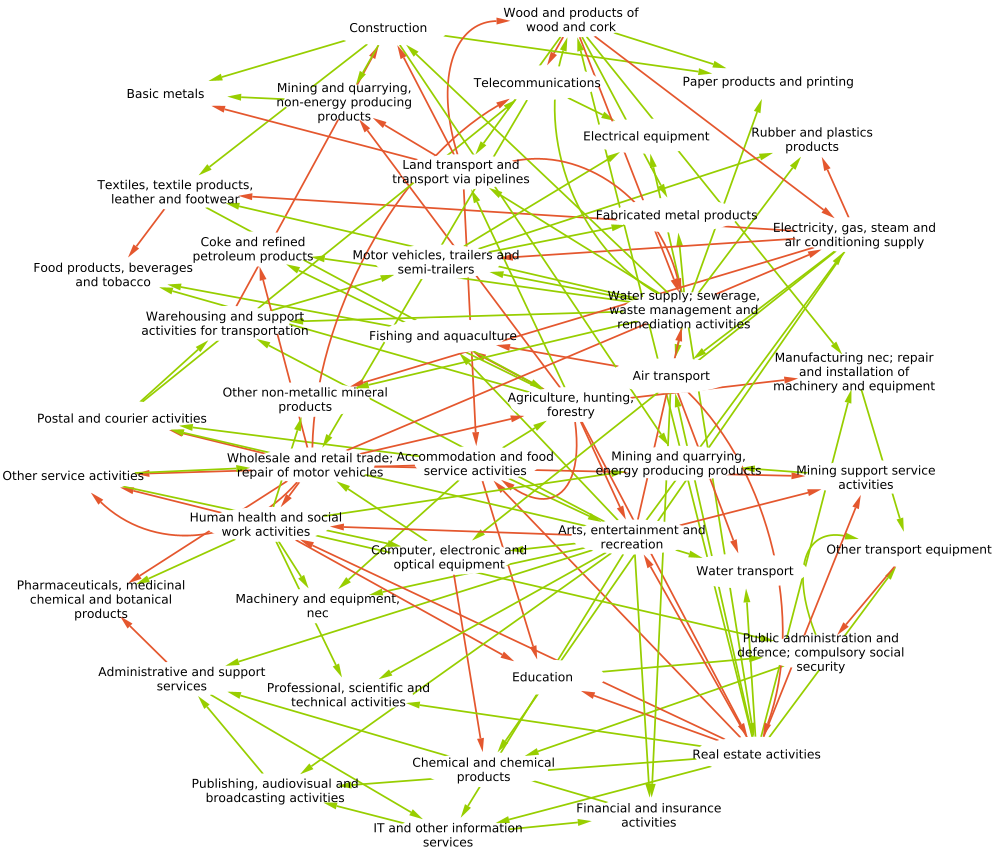}}
    \caption{Resultant value chain network that consists of nodes representing industries and directed edges representing the input-output connections between the industries. The edges are labelled using the OECD labelled data, with red edges indicating misclassified links and green edges indicating correctly identified inputs by the Backward \& Forward method.}
    \label{fig:oecd_value_chains_result}
    \end{center}
    \end{figure}

    In Figure \ref{fig:oecd_value_chains_result} we can see the value chain network produced by the ``Backward \& Forward'' model using the best parameters. With the OECD labeled data, we are able to colour the correct links with green and the incorrect ones with red. There are 84 true-positive and 48 false-positive input-output links.
    
    \section{Results}
    To produce our final results, we apply the ``Forward \& Backward'' method on every traded product to get the 3 top input candidates. For the model's parameters we use $rca\_locations\_1 = 2$, $rca\_industries\_1 = 1.5$, $rca\_locations\_2 = 3.5$  and $rca\_industries\_2 = 2$.\\

\begin{figure}[ht]
    \begin{center}    \centerline{\includegraphics[width=1\columnwidth]{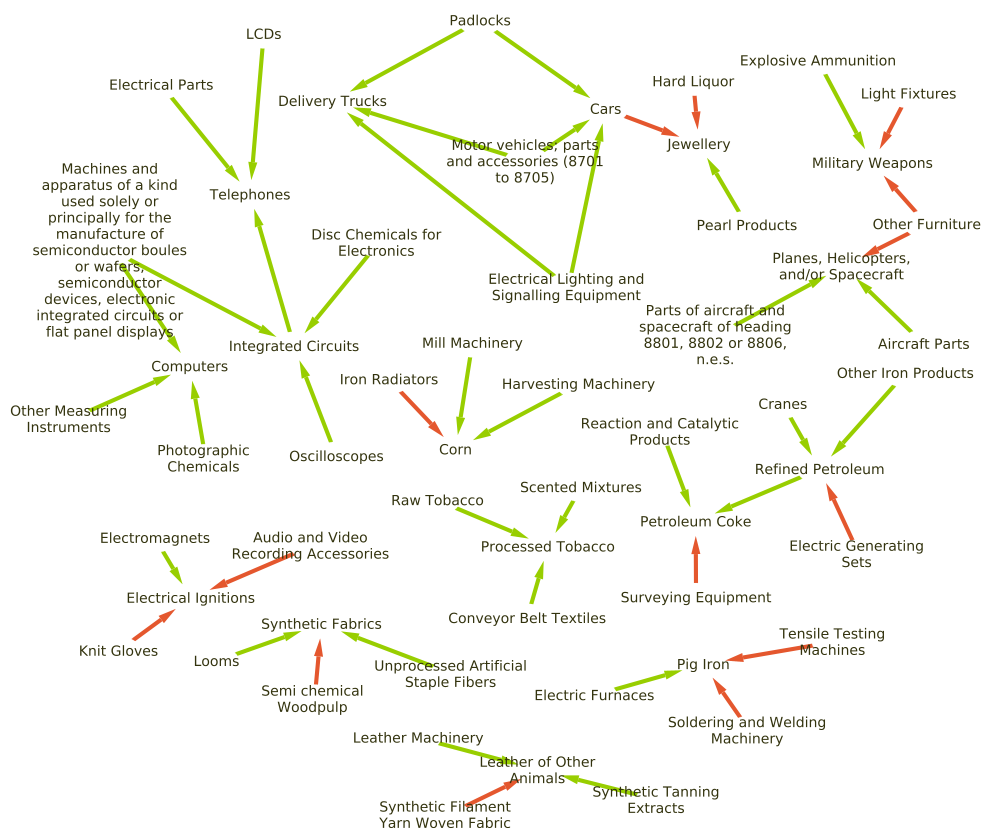}}
    \caption{A subset of our resultant value chain network that consists of nodes representing HS4 products and directed edges representing the input-output connections between the products. The edges are manually labelled, with red edges indicating misclassified links and green edges indicating correctly identified inputs.}
    \label{fig:oec_value_chain}
    \end{center}
    \end{figure}

  Our initial application of the ``Backward \& Forward'' method focuses on 1200 HS4 products. In Figure \ref{fig:oec_value_chain} we see part of the value chain networks produced by the ``Backward \& Forward'' method using HS4 trade data. These examples were validated manually. Red edges represent false positive, while the green edges are true positive value chain relationships.\\ 
     Examples of accurately identified products are the inputs of ``Cars'' and ``Delivery Trucks'' where for both we get ``Motor vehicle parts and accessories (8701 to 8705)'', ``Electrical Lighting and Signaling Equipment'' and ``Padlocks''. Other examples are ``Telephones'' and ``Computers'' where the inputs for the former are ``LCDs'', ``Electrical Parts'' and ``Integrated Circuits'', and for the latter are ``Photographic Chemicals'', ``Machines and apparatus of a kind used solely or principally for the manufacture of semiconductor boules or wafers, semiconductor devices, electronic integrated circuits or flat panel displays'' and ``Other Measuring Instruments''. The other correct results we can see in the figure are for the products ``Processed Tobacco'' and ``Integrated Circuits''. \\
     However, we also see some false positive results in Figure \ref{fig:oec_value_chain}. In the example of ``Electrical Ignitions'' the model incorrectly predicts as inputs ``Audio and Video Recording Accessories'' and ``Knit Gloves'' while it correctly predicts ``Electromagnets''. Other examples where our model is able to identify only one correct input are ``Military Weapons'', ``Jewellery'' and ``Pig Iron''.

    \begin{table}
    \centering
    \small
    \setlength\tabcolsep{2pt}
    \resizebox{\columnwidth}{!}{%
        \begin{tabular}{|l|rrr|}
        \hline
        Output & Input & Input & Input\\
        \hline
    Cars & \cellcolor{green!30}Motor vehicles parts & \cellcolor{green!30}Electrical Lighting/Signalling  & \cellcolor{green!30}Padlocks \\
    Delivery Trucks & \cellcolor{green!30}Motor vehicles parts & \cellcolor{green!30} Electrical Lighting/Signalling & \cellcolor{green!30} Padlocks \\
    Processed Tobacco & \cellcolor{green!30} Raw Tobacco & \cellcolor{green!30} Scented Mixtures & \cellcolor{green!30} Conveyor Belt Textiles \\
    Integrated Circuits & \cellcolor{green!30}Chemicals for Electronics & \cellcolor{green!30} Apparatus for semiconductors & \cellcolor{green!30} Oscilloscopes \\
    Telephones & \cellcolor{green!30} Integrated Circuits & \cellcolor{green!30}Electrical Parts & \cellcolor{green!30}LCDs \\
    Computers & \cellcolor{green!30} Photographic Chemicals & \cellcolor{green!30} Other Measuring Instruments & \cellcolor{green!30} Apparatus for semiconductors\\
    Petroleum Coke & \cellcolor{green!30} Refined Petroleum & \cellcolor{green!30} Surveying Equipment & \cellcolor{green!30} Reaction and Catalytic Products \\
    Refined Petroleum & \cellcolor{green!30} Other Iron Products & \cellcolor{red!30} 
     Electric Generating Sets & \cellcolor{green!30} Cranes \\
     Electrical Ignitions & \cellcolor{red!30}  Knit Gloves & \cellcolor{green!30} Electromagnets & \cellcolor{red!30} Audio-Video Recording \\
    Leather of Animals & \cellcolor{green!30} Leather Machinery & \cellcolor{green!30} Synthetic Tanning Extracts & \cellcolor{red!30} Synth. Filam. Yarn Woven Fabric \\
    Synthetic Fabrics & \cellcolor{green!30} Looms & \cellcolor{green!30} Unprocessed Artificial Staple Fibers & \cellcolor{red!30}  Semi chemical Woodpulp \\
    Corn & \cellcolor{green!30} Mill Machinery & \cellcolor{red!30}  Iron Radiators & \cellcolor{green!30} Harvesting Machinery \\
    Jewellery & \cellcolor{red!30}  Cars & \cellcolor{red!30} Hard Liquor & \cellcolor{green!30} Pearl Products \\
    Pig Iron & \cellcolor{green!30} Electric Furnaces & \cellcolor{red!30} Tensile Testing Machines & \cellcolor{red!30} 
 Soldering/welding Machinery \\
    Plane, Helicop., Spacecraft & \cellcolor{green!30} Aircraft Parts & \cellcolor{green!30} Parts of aircraft and spacecraft & \cellcolor{red!30}  Other Furniture \\
    Military Weapons & \cellcolor{red!30} 
 Other Furniture & \cellcolor{green!30} Explosive Ammunition & \cellcolor{red!30} Light Fixtures \\
        \hline
        \end{tabular}
        }
    \caption{HS4 examples produced by the Backward \& Forward method where the green cell represents a correctly predicted input candidate and red incorrectly. For readability, some of the product names have been shortened.}
    \label{tab:hs4_examples}
    \end{table}

\begin{figure*}
    \begin{center}
\centerline{\includegraphics[width=1\linewidth]{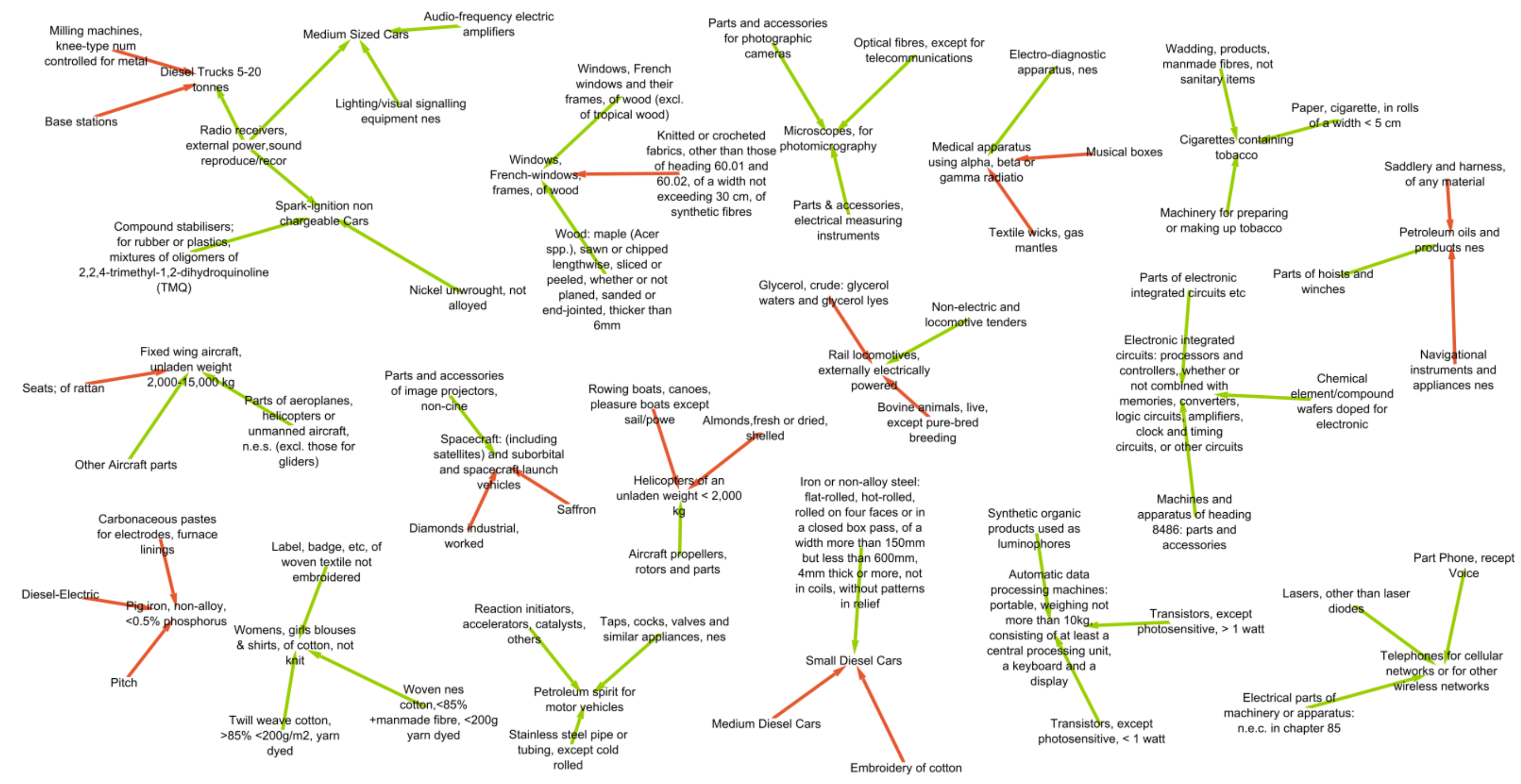}}
    \caption{ A subset of our resultant value chain network that consists of nodes representing HS6 products and directed edges representing the input-output connections between the products. The edges are manually labelled, with red edges indicating misclassified links and
    green edges indicating correctly identified inputs.}
    \label{fig:hs6_result_network}
    \end{center}
    \end{figure*}

    Next, our method is applied to a more granular product classification, namely the HS6, which consists of over 5000+ distinct products. In Figure \ref{fig:hs6_result_network} we see examples of the value chain networks produced by the ``Backward \& Forward'' method using HS6 products. Notably, our method accurately identifies the inputs of ``Medium Sized Cars'', ``Cigarettes containing tobacco'', ``Telephones for cellular networks or for other wireless networks'' and ``Electronic integrated circuits: processors and controllers, whether or not combined with memories, converters, logic circuits, amplifiers, clock and timing circuits, or other circuits'' even at this more granular level of product classification. \\
     The model predicted false positive results for ``Helicopters of an unladen weight $<$ 2,000 kg'' with the inputs ``Rowing boats, canoes, pleasure boats except sail/powe'' and ``Almonds,fresh or dried, shelled''. The product ``Pig iron, non-alloy, $<$0.5\% phosphorus'' has no correct inputs.

    \subsection{Validation}
    To evaluate our estimation of the $L_{p_1p_2}$ term using the ``Backward \& Forward'' method we compare it with a baseline model which randomly picks out three inputs for each product.
    Due to the lack of a benchmark to validate the trade data results produced by the ``Backward \& Forward'', we perform a random sampling of 50 products. Then we manually label the identified inputs of those 50 products. 
    
\begin{figure}[h]
    \begin{center}    \centerline{\includegraphics[width=\columnwidth]{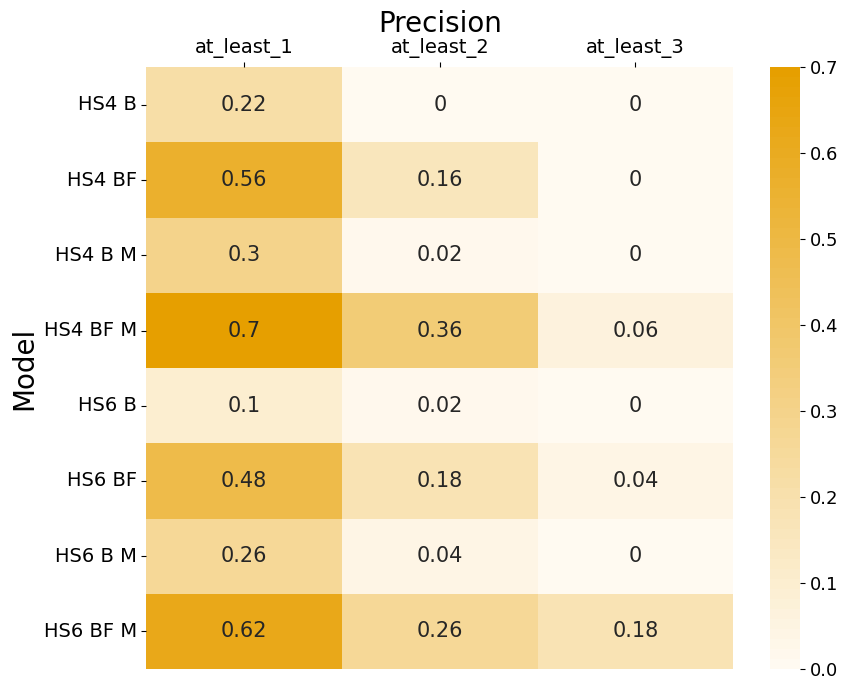}}
    \caption{Heatmap showing our manually validated results by randomly sampling 50 input-output relationships. The values represent the percentage of the 50 examples that in the first column had at least one correctly identified input product, in the second column at least two and in the third column all of the input products identified correctly by the Backward \& Forward model using the different types of models (B -- Baseline and BF -- Backward \& Forward) and data (HS4, HS6 and M -- Machinery products) specified on the y-axis.}
    \label{fig:validation}
    \end{center}
    \end{figure}

    Figure \ref{fig:validation} shows the percentage of outputs having at least one, two and three inputs correctly identified. On the y-axis we have the different models: ``Baseline Model'' (B) and ``Backward \& Forward Model'' (BF) using the HS6 and HS4 product classifications. Furthermore, we test the case where the output products are only part of the ``Machinery" group (M). In the Figure \ref{fig:validation} we see that our model is able to identify at least one of the inputs for 70\% of the 50 HS4 products coming from the group of ``Machinery", whereas the baseline is able to identify only 30\%. We can see similar performance between the ``Baseline" and the ``Backward \& Forward Model" in the other cases as well. With this, we conclude that the ``Backward \& Forward" performs significantly better than the ``Baseline Model".
    
\section{Conclusion}
Here we presented a first attempt to learn value chain relationships and estimate trade flows from trade data, by combining concepts from trade theory with machine learning techniques. While data on global value chains is notoriously aggregated, the "Forward \& Backward" method offers a promising solution for mapping global value chains at the product level.\\

However, it is important to acknowledge that our method is not perfect. Although it operates at the product level, it may provide some false-positive value chain relationships, and it does not offer a complete input-output network. Additionally, optimizing the four different parameters of our method can be a slow and complicated process.\\

Despite these limitations, the ``Backward \& Forward'' method outperforms the random baseline, identifying at least one correct input for more than 40\% of the data, which is more than twice as much as the capabilities of a random baseline model. We also demonstrate that the method identifies correctly three inputs for complex products such as cars, integrated circuits, computers, and telephones. This validates the possibility of using international trade data at the regional level to identify value chain relationships.\\

Increasing the accuracy of the ``Backward \& Forward'' method represents an interesting avenue for future research. One approach is by fine-tuning the model with input-output tables that have a higher sectoral and geographical resolution than the present OECD  ICIO data. Another approach is to expand the regional trade data by linking product codes with different classifications (e.g. SITC, CPC, SIC, GTAP) to HS. Lastly, extending the model to predict more than three inputs per output would bring us closer to obtaining a complete value chain network.

\ack We thank Ron Boschma, the participants of the International PhD course in Economic Geography, Utrecht 2022, and the members of the Center for Collective Learning for their valuable input. This project was supported by the National Institute of Standards and Technology, ANR-19-PI3A-0004 and LearnData ERA Chair Grant agreement ID: 101086712.

\bibliography{ecai}

\section{Appendix}
\subsection{OEC Trade Data Manipulation}
We use regional trade data from the Observatory of Economic Complexity (oec.world) for our results.\\
We combine the data for the 306 regions into a table containing as columns: ``geography''-- name of the region, ``product''-- HS4 product name, ``value\_imp''-- the import trade value, ``value\_exp''-- the export trade value, ``geography\_exp''-- total exports of the region, ``geography\_imp''-- total imports of the region, ``product\_exp''-- total exports of the product, ``product\_imp''-- total imports of the product, and two classical measures of specialization ``rca\_exp'' and ``rca\_imp.''

$$RCA_{exp}=\frac{\frac{value\_exp}{geography\_exp}}{\frac{product\_exp}{sum(value\_exp)}},$$ 
    
$$RCA_{imp}=\frac{\frac{value\_imp}{geography\_imp}}{\frac{product\_imp}{sum(value\_imp)}}$$
    
In the Table \ref{tab:oec_table_final} we can see the final look of our trade data.

    \begin{table*}
    \centering
    \small
    \setlength\tabcolsep{2pt}
        \resizebox{0.9\linewidth}{!}{%
        \begin{tabular}{|lrrrrrrrrrr|}
        \hline
        {} &         geography &  value\_imp &                   product &  geography\_imp &   product\_imp &   rca\_imp &   value\_exp &  geography\_exp &   product\_exp &     rca\_exp \\
        \hline
        0 &  Ciudad de México &  5882480.0 &  Other Vegetable Products &   4.101108e+11 &  6.901484e+10 &  0.011240 &   1775336.0 &   3.633443e+11 &  1.509038e+09 &    1.057660 \\
        1 &        Guanajuato &  2468431.0 &  Other Vegetable Products &   3.754898e+10 &  6.901484e+10 &  0.051516 &   4320249.0 &   3.447602e+10 &  1.509038e+09 &   27.125367 \\
        2 &           Jalisco &  9451035.0 &  Other Vegetable Products &   9.338915e+10 &  6.901484e+10 &  0.079306 &  34593964.0 &   6.327343e+10 &  1.509038e+09 &  118.348546 \\
        3 &  Estado de México &   508352.0 &  Other Vegetable Products &   6.436369e+10 &  6.901484e+10 &  0.006189 &   2616971.0 &   3.181877e+10 &  1.509038e+09 &   17.803261 \\
        4 &        Nuevo León &  3813095.0 &  Other Vegetable Products &   1.256306e+11 &  6.901484e+10 &  0.023785 &  51550913.0 &   1.262054e+11 &  1.509038e+09 &   88.418374 \\
         ... &  ... & ... & ... &   ... &  ... &  ... &  ... &   ... &  ... &   ... \\
         196117 &  Metropolitana De Santiago &    6178.30 &         Titanium Ore &   2.233049e+11 &  1.166828e+11 &  0.000013 &    551658.00 &   2.108547e+11 &  2.649914e+08 &   3.225070 \\
        196118 &  Metropolitana De Santiago &     172.95 &      Granulated Slag &   2.233049e+11 &  3.079868e+08 &  0.000136 &    335812.32 &   2.108547e+11 &  4.000072e+10 &   0.013006 \\
        196119 &  Metropolitana De Santiago &  848039.05 &  Prepared Explosives &   2.233049e+11 &  1.898840e+09 &  0.108167 &  38389897.04 &   2.108547e+11 &  1.352253e+09 &  43.980460 \\
        196120 &  Metropolitana De Santiago &  483473.53 &        Railroad Ties &   2.233049e+11 &  4.331550e+09 &  0.027033 &    191154.59 &   2.108547e+11 &  7.041212e+08 &   0.420570 \\
        196121 &  Metropolitana De Santiago &     274.74 &      Scrap Aluminium &   2.233049e+11 &  5.365896e+10 &  0.000001 &  98316711.65 &   2.108547e+11 &  1.346654e+11 &   1.131025 \\
        \hline
        \end{tabular}%
        }
    \caption{OEC Final Trade Data Table}
    \label{tab:oec_table_final}
    \end{table*}

   \subsection{OECD Data Manipulation}
    To fine-tune our ``Backward and Forward'' method we use the OECD Inter-Country Input-Output (ICIO) Tables.
    Here, we work with the intermediate use at basic prices data from 2016, 2017 and 2018, and we merge them by summing up the trade flows. To clean the data, we merge the data of China ``CN1'', ``CN2'' and ``CHN'' into "CHN". We do the same to the Mexico data, ``MX1", ``MX2" and ``MEX" into ``MEX". We also remove the country ``ROW" which represents ``Rest of the World" as it can introduce noise to the model.\\
    
    Concerning the industries, we remove the industry 97T98 ``Activities of households as employers; undifferentiated goods- and services-producing activities of households for own use" as it has not been traded.\\
    We end up with a matrix of 2904x2904 industries-country pairs (44x66).\\
    Moreover, we put the trade flow of reexports, export from and to the same country, to 0.\\
    
    First, we start by creating the tunning table. For every country-industry pair, we put the country code in "geography" column and the product code in "product" column. Then in ``value\_imp" column we sum over the columns and for the ``value\_exp" we sum over the rows. Next, we calculate the ``geography\_exp" by grouping over ``geography" and summing up the ``value\_exp", and ``product\_exp" by grouping over "product" and summing up the ``value\_exp". We do the same to calculate the ``geography\_imp", by grouping over ``geography" and summing up the ``value\_imp", and ``product\_imp" by grouping over ``product" and summing up the ``value\_imp". As we did for the Trade Data, using the formula $\frac{\frac{value\_exp}{geography\_exp}}{\frac{product\_exp}{sum(value_exp)}}$ we calculate ``rca\_exp". And we use the formula  $\frac{\frac{value\_imp}{geography\_imp}}{\frac{product\_imp}{sum(value_imp)}}$ to calculate ``rca\_imp".
    \\Table \ref{tab:oecd_table_final} shows the final look of the training data.
    
    \begin{table*}
    \centering
    \small
    \setlength\tabcolsep{2pt}
        \resizebox{0.8\linewidth}{!}{%
        \begin{tabular}{|rrrrrrrrrrr|}
        \hline
        {}  &     value\_imp & geography & product &  geography\_imp &   product\_imp &   rca\_imp &     value\_exp &  geography\_exp &    product\_exp &   rca\_exp \\
        \hline
        0  &  11619.407384 &       AUS &   01T02 &  353818.854850 &  835032.99528 &  1.440590 &  22935.551244 &  595965.513523 &  768630.020805 &  1.834052 \\
        1 &   4840.318308 &       AUT &   01T02 &  321604.777528 &  835032.99528 &  0.660220 &   2193.216844 &  349635.801449 &  768630.020805 &  0.298944 \\
        2 &   7049.627146 &       BEL &   01T02 &  526307.471482 &  835032.99528 &  0.587576 &   4927.290870 &  532869.434581 &  768630.020805 &  0.440667 \\
        3  &  27917.032198 &       CAN &   01T02 &  876406.697376 &  835032.99528 &  1.397336 &  37023.829481 &  909091.443344 &  768630.020805 &  1.940874 \\
        4 &   3671.147871 &       CHL &   01T02 &   93649.791607 &  835032.99528 &  1.719620 &   7221.641047 &  174677.439401 &  768630.020805 &  1.970256 \\
        ... &   ... &       ... &   ... &   ... &  ... &  ... &   ... &  ... &  ... &  ... \\
        2899  &   521.915500 &       ZAF &   94T96 &  147558.428838 &  253890.016018 &  0.510307 &   1.420904 &  158225.011733 &  15681.17867 &  0.020977 \\
        2900  &  2375.654280 &       TWN &   94T96 &  576254.654859 &  253890.016018 &  0.594791 &  93.548425 &  693550.512370 &  15681.17867 &  0.315080 \\
        2901  &  4563.273980 &       THA &   94T96 &  473645.715393 &  253890.016018 &  1.390011 &   0.000000 &  396945.386345 &  15681.17867 &  0.000000 \\
        2902 &    41.091241 &       TUN &   94T96 &   29409.824719 &  253890.016018 &  0.201582 &   6.295216 &   21764.013076 &  15681.17867 &  0.675669 \\
        2903  &  1962.493248 &       VNM &   94T96 &  502262.634668 &  253890.016018 &  0.563732 &   1.248322 &  345511.049333 &  15681.17867 &  0.008440 \\
        \hline
        \end{tabular} %
        }
    \caption{Final OECD Table}
    \label{tab:oecd_table_final}
    \end{table*}
    
    Then, we move on to creating the labeled data. For this, we take the initial table (country-industry x country-industry pairs and we sum over the industries so that we get a matrix of 44x44 industries with their corresponding trade flows. \\
    Afterwards, we apply the trade intensity formula for every input-output industry and get the Table \ref{tab:trade_intensty_binary_table}. 
    Trade Intensity (TI). TI index, according to Yamazawa\cite{yamazawa1970intensity}, is used to determine whether the value of trade between two countries is greater or smaller than would be expected on the basis of their importance in world trade. It is defined as the share of one country’s exports going to a partner divided by the share of world exports going to the partner. \\
    In our situation we will be using the TI between products (industries).
    
    We calculate it as:
    \begin{equation}
    TI_{pp'} = \frac{X_{pp'}/\sum_{p'}X_{pp'}}{\sum_{p}X_{pp'}/\sum_{pp'}X_{pp'}}
    \end{equation}
    
    Where $X_{pp'}$ and $\sum_{p}X_{pp'}$ are the trade flows from industry $p$’s and from all the industries to industry $p'$, and where $\sum_{p'}X_{pp'}$ and $\sum_{pp'}X_{pp'}$ are industry $p$’s total exports and total export of all industries respectively. An index of more (less) than one indicates a trade flow that is larger (smaller) than expected between the two products (industries).\\
    
    Then, we binarize the matrix by applying 
    $\begin{dcases}
    1,& \text{if } TI \geq 1\\
    0,              & \text{otherwise}
    \end{dcases}$.\\
    The labeled data that we get is depicted in Table \ref{tab:trade_intensty_binary_table}. \\

    When we map the entire input-output network between the industries we get the value chain shown in the Figure \ref{fig:oecd_value_chains}.

    \begin{table*}[ht]
    \centering
    \small
    \setlength\linewidth{1pt}
        \begin{tabular}{|l|rrrrrrrrrrr|}
        \hline
        {} &  01T02 &  03 &  05T06 &  07T08 &  09  & ... & 
        84 &  85 &  86T88 &  90T93 &  94T96\\
        \hline
        01T02 &      1 &   1 &      0 &      0 &   0 & ... &
        1 &   1 &      1 &      1 &      1 \\
        03    &      0 &   1 &      0 &      0 &   0  & ... &
           1 &   1 &      1 &      1 &      1 \\
        05T06 &      0 &   1 &      1 &      1 &   1 & ... &
           1 &   1 &      1 &      1 &      1\\
        ... &  ... &   ... &  ... &  ... &  ...  & ... & 
        ... &  ... &   ... &  ... &  ...  \\
        86T88 &      0 &   0 &      0 &      0 &   0 & ... & 
         0 &   1 &      1 &      0 &      1 \\
        90T93 &      0 &   0 &      0 &      0 &   0 & ... & 
         1 &   1 &      1 &      1 &      1\\
        94T96 &      0 &   0 &      0 &      0 &   0 & ... & 
         0 &   1 &      0 &      0 &      1 \\
        \hline
        \end{tabular}
    \caption{Trade Intensity Binary OECD Table}
    \label{tab:trade_intensty_binary_table}
    \end{table*}

    \begin{figure*}[htb]
    \begin{center}
\centerline{\includegraphics[width=0.8\linewidth]{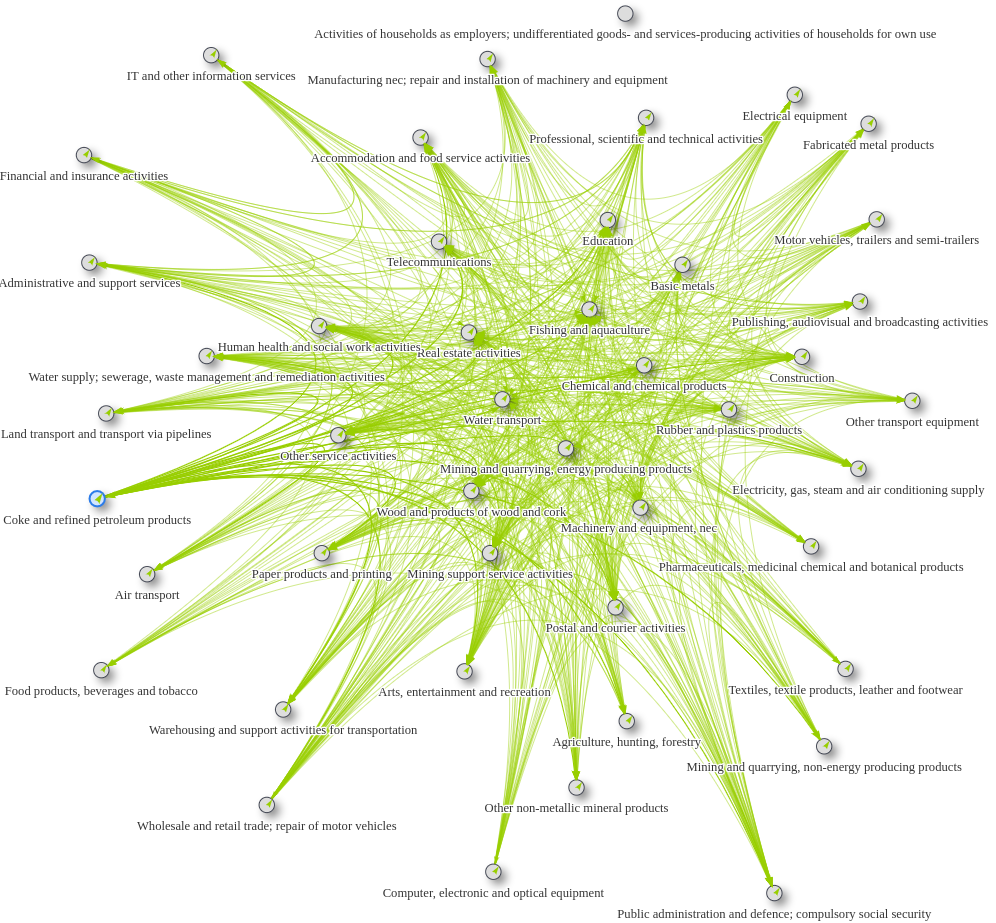}}
    \caption{Depicts the full OECD binarized/labeled network that consists of nodes representing sectors and edges representing the input-output connections between the sectors. }
    \label{fig:oecd_value_chains}
    \end{center}
    \end{figure*}

\end{document}